%% file: qfast.tex
\documentclass[runningheads]{llncs}

\usepackage{color,verbatim,afterpage,url,xfrac}
\graphicspath{{figures/}}
 \long\def\comment#1{}

\def\parah#1{\vspace*{0.0in} \noindent{\bf #1:}}

\usepackage{algorithm}
\usepackage{algorithmicx}
\usepackage{algpseudocode}

\usepackage{amsmath}
\usepackage{amsfonts}
\usepackage{braket}
\usepackage{graphicx}
\usepackage{braket}

\newcommand{\I}{\mathrm{i}\mkern1mu}

\usepackage{amssymb}

\definecolor{klcolor}{rgb}{0.5,0.7,0.9}
\definecolor{eycolor}{rgb}{0.05,0.4,0.1}
\definecolor{kscolor}{rgb}{0.9,0.1,0.1}

\begin{document}

\title{QFAST: Quantum Synthesis Using a Hierarchical Continuous Circuit Space}

\author{Ed Younis\inst{1,2} \and
Koushik Sen\inst{1} \and Katherine Yelick\inst{1,2} \and Costin Iancu\inst{2}}

\authorrunning{Younis et al.}

\institute{University of California Berkeley \and Lawrence Berkeley National Laboratory}

\maketitle

\begin{abstract}
 {\it  We present QFAST, a quantum synthesis tool designed to
  produce short circuits and to scale well in practice. Our contributions
  are: 1) a novel representation of circuits able to encode placement
  and topology; 2) a hierarchical approach with an iterative
  refinement 
  formulation that combines ``coarse-grained'' fast optimization during
  circuit structure search with a good, but slower, optimization stage only
  in the final circuit instantiation stage.  When compared against
  state-of-the-art techniques, although not optimal,  QFAST can generate
  much shorter circuits for ``time dependent evolution'' algorithms used by domain scientists. We also show the composability and
  tunability of our formulation in terms of circuit depth and running
  time. For example, we show how to generate shorter circuits by
  plugging in the best available third party synthesis algorithm at a
  given hierarchy level. Composability enables portability across chip
architectures, which is missing from the available  approaches. }
  
\end{abstract}

\input{sections/introduction.tex}

\input{sections/background.tex}
\input{sections/circuit.tex}
\input{sections/algorithm2.tex}
\input{sections/props}

\input{sections/eval.tex}
\input{sections/discussion.tex}

\input{related}

\input{sections/conc}

\section{Acknowledgment}
This work was supported by the DOE under contract DE-5AC02-05CH11231, through the Office of Advanced Scientific Computing Research (ASCR) Quantum Algorithms Team and Accelerated Research in Quantum Computing programs.

\bibliographystyle{splncs04}
\bibliography{bibliography,quantum,quant_chem}

\begin{appendix}
\input{sections/psuedocode}
\end{appendix}

\end{document}

%% file: sections/introduction.tex
\section{Introduction} \label{intro}

Quantum computing has the potential to provide transformational societal impact at the decade threshold. 
 As quantum programming is subtle and with a very steep learning curve,  one of the important prerequisites for success is the ability to generate programs from high level problem descriptions. Quantum synthesis (or compilation\footnote{Originally synthesis was referred to as quantum compiling within the Quantum Information Science community.}) is perhaps the most powerful approach available to assist in algorithm discovery, hardware exploration or quantum program optimization. Ideally for adoption, synthesis will need to generate short circuits fast, in a hardware/topology specialized manner. Synthesis has a distinguished history~\cite{iten2016quantum,shende2006synthesis,tucci2005introduction,DawsonNielson05,ZXZ16,BocharovPRL12,MIM13,Qcompile16,ctmq,23gates,householderQ,CSD04,amy16,seroussi80} but practical adoption has been hampered by perceived shortcomings in most requirements: 1) generated circuits are long; 2) algorithms are slow; and 3) techniques are not topology-aware, hence generate long circuits or are hard to specialize for a different gate set. In this work we present a tunable synthesis approach able to generate reasonably short circuits in time acceptable for  practical purposes: {\it our design metrics are circuit quality and speed to solution}. In order to make synthesis usable and to enable scientific discovery, we aim to generate circuits that are shorter than those produced by state-of-the-art fast techniques~\cite{synthcsd,tucci,iten2016quantum,uq} while  closer to the depth generated by optimal  slow techniques. 

Currently, an executable quantum program is described by a circuit as a space-time evolution of gates/operators on qubits/wires. This model of computation is likely to survive for the foreseeable future. Synthesis takes as input a high level description of the computation as a unitary matrix and produces a circuit executable on hardware. As programs are circuits that use hardware resources, the first goal of synthesis is to minimize resource consumption, equated with the total number of gates or circuit depth. This is true  long term, but even more important in the current (and near-future) stage where we deploy Noisy Intermmediate-Scale Quantum devices. NISQ devices are characterized by high error rates, in particular on multi-qubit operations, and the general expectation is that running meaningful algorithms will require a painstaiking depth optimization process to eliminate multi-qubit operations.  For existing superconducting architectures with two-qubit {\it CNOT} gates,  {\it our first optimality target is minimizing} their {\it count in the generated circuit}. This metric is exhaustively~\cite{qaqc,raban,synthcsd,noisemap} used by other existing work. In particular, Davis et al~\cite{davis2019heuristics} very recently introduced a technique able to generate minimal length circuits in a topology-aware manner, but they do so at the expense of running time. Their approach gives us a first threshold: we aim to generate  circuits faster while close to optimal depth.

{\it The second design criteria for our approach is speed}: we aim  to provide a solution within an acceptable and usable time interval.  To our knowledge, the fastest existing techniques are based on linear algebra matrix decomposition as illustrated by the work of Iten et al~\cite{shende2006synthesis,iten2016quantum,uq}. This gives us a second threshold: we want to generate circuits shorter than theirs.

Intuitively, our Quantum Fast Approximate Synthesis Tool (QFAST) succeeds by embracing and combining  the strengths behind the design principles of these state-of-the-art synthesis techniques. Fast algorithms  employ coarse grained multi-qubit fixed function building blocks. The only optimal approach~\cite{davis2019heuristics} known to work at three qubits or more uses  continuous  representations of hardware native gates and combines numerical optimization with the proven optimal A*  search algorithm. In its attempt to reach optimal depth, QFAST uses a continuous representation of multi-qubit general operators and numerical optimization.  In its attempt to run fast, QFAST  tunes the  operator granularity in qubits and instead of combinatorial search it performs a single combined step  of structural and functional optimization.

For a $n$ qubit unitary, the algorithm starts by trying to determine the structure of a circuit that uses $m < n$ generic qubit operators using numerical optimization.  The optimization criteria is the ``distance'' between the solution and the original unitary matrix. The first stage is {\it expansion} where the circuit is grown layer by layer by one $m$ block.
At each expansion stage,  we use first coarse-grained optimization called {\it exploration} to determine block  placements on qubits, followed by fine-grained optimization called {\it refinement} to finalize the functions computed by each block. After building a circuit using $m$ qubit blocks, we expand each block into finer grained blocks.
This stops when we reach two qubit generic gates, where we apply optimal KAK~\cite{kak} decomposition. 

The main contributions of QFAST 
are:
\begin{enumerate}
\item A novel representation of multi-qubit circuits able to encode placement and topology.
\item  A hierarchical approach with a iterative refinement 
  formulation that combines ``coarse-grained'' fast optimization during
  circuit structure search with a good, but slower, optimization stage only
  in the final instantiation stage.
\item A composable, retargetable and tunable methdology able to exploit third party synthesis algorithms at the qubit granularity deemed necessary for depth optimality or speed purposes. 
\end{enumerate}

QFAST has been evaluated on a collection of circuits including depth optimal~\cite{noisemap} circuits, fixed lengh parameterized circuits that appear in VQE~\cite{McClean2015} and QAOA~\cite{farhi2014quantum} formulations and circuits for time dependent Hamiltonians~\cite{tfimshin,tfimlb} (TFIM). The results indicate that while sub-optimal, QFAST scales much better than the optimal synthesis formulation.  When compared
  directly with the
  state-of-the-art UniversalQ~\cite{uq}
  fast approach based on numerical decomposition, QFAST is slower but
  can generate circuits that are shorter by a factor of $5.7 \times$ on
  average and up to $46.7 \times$. We also show the composability and
  tunability of our formulation in terms of circuit depth and running
  time. For example, we can plug in at any step of expansion the best known optimizer for the given granularity.

 Overall we find these results to be  very promising and to bode well for the future adoption of synthesis in the quantum software development toolkit. In particular, none of the existing solutions, either synthesis or optimizing compilers, reduce the depth of VQE and TFIM circuits. QFAST was able to reduce their depth by a factor of $6.3 \times$ on average and up to $30 \times$.  
 QFAST provides a practical and tunable approach that generates short enough circuits in an acceptable amount of time. The composability enables easy retargeting to architectures with different gate sets.  It is enough to plug in the specialized synthesis module for small scale, such as a KAK implementation for the given target.The scalability of our method is likely to be sufficient for practical impact  within the NISQ era  forecast.

The rest of this paper is structured as follows. 
In the next section, we review the necessary background on quantum computation. In section \ref{sec:circuit} we introduce our novel continuous structure of the circuit space, which we use in the section \ref{sec:alg} to build and analyze a synthesis algorithm. We include an in-depth evaluation of this method compared to both the UniversalQ and Search Compilers in sections \ref{sec:eval}. We end with a discussion in section \ref{disc} and comment on related works in section \ref{sec:related}.

%% file: sections/background.tex
\section{Background} \label{background}

 A qubit is an element of the Hilbert space $\mathbb{C}^2$ of 2-dimensional complex vectors. Typically, a qubit's state is represented in Dirac's notation $\ket{\psi}$ which is a column-vector $\begin{pmatrix}a_0 \\ a_1\end{pmatrix}$ of $\mathbb{C}^2$. We refer to the basis states as $\ket{0} = \begin{pmatrix}1 \\ 0\end{pmatrix}$ and $\ket{1} = \begin{pmatrix}0 \\ 1\end{pmatrix}$. The qubit state $\ket{\psi} = \begin{pmatrix}a_0 \\ a_1\end{pmatrix}$ can be represented as $\ket{\psi} = a_0 \ket{0} + a_1 \ket{1}$ using the basis states.  We can join multiple qubit's state into one quantum system with an outer product or tensor product of the states of the individual qubits. For example, the three qubit state $\ket{\psi}$ resulting from joining the qubits $\ket{\psi_0}$, $\ket{\psi_1}$ and $\ket{\psi_2}$ is $\ket{\psi_0} \otimes \ket{\psi_1} \otimes \ket{\psi_2}$ or equivalently $\ket{\psi_0\psi_1\psi_2}$. When context is clear we will refer to multiple qubit states simply by $\ket{\psi}$. 
It follows that the state space for an n-qubit system is $\mathbb{C}^{2^n}$. A \emph{pure state} is a state $\ket{\psi} = \begin{pmatrix}a_0 & a_1 & \hdots & a_{2^{n}-1}\end{pmatrix}^{T}$ that satisfies the constraint $\sum_i^{2^n}{|a_i|^2} = 1$.  Quantum programs operate on pure states; in the rest of the paper will use the term state to mean a pure state.

\parah{Quantum Operators} Quantum operators transform a state $\ket{\psi}$ to another state $\ket{\psi'}$. Each such operator could be denoted by a unitary $2^n \times 2^n$ matrix, where $n$ is the number of qubits that the operator takes as input. Note that a matrix $U$ is unitary if it's conjugate transpose $U^\dagger$ is it's inverse, i.e. $UU^\dagger = U^\dagger U = I$. Some basic quantum operators are often referred to as {\it{gates}}. The application of a quantum operator $U$ on a quantum state is denoted by $U\ket{\psi}$. A few examples of common operator are $X, Y, Z, {\rm\it CNOT}$ whose corresponding unitary matrices are the following:
{\footnotesize
$$
\ X={\begin{pmatrix}0&1\\1&0\end{pmatrix}}\ \ 
\ Y={\begin{pmatrix}0&-i\\i&0\end{pmatrix}}\ \ 
\ Z={\begin{pmatrix}1&0\\0&-1\end{pmatrix}}\ \
\ {\rm\it CNOT}={\begin{pmatrix}1&0&0&0\\0&1&0&0\\0&0&0&1\\0&0&1&0\end{pmatrix}}\ \
$$
}
A couple of examples applying a gate on a concrete state and the resulting state is shown below:
{\footnotesize
$$
X\ket{0} = \begin{pmatrix}0&1\\1&0\end{pmatrix} . \begin{pmatrix}1 \\ 0\end{pmatrix} = \begin{pmatrix}0 \\ 1\end{pmatrix} = \ket{1}
$$
$$
X(a_0\ket{0} + a_1\ket{1}) = \begin{pmatrix}0&1\\1&0\end{pmatrix} . \begin{pmatrix}a_0 \\ a_1\end{pmatrix} = \begin{pmatrix}a_1 \\ a_0\end{pmatrix} = a_0 \ket{1} + a_1 \ket{0}
$$
}
\noindent The X, Y, and Z gates are single qubit Pauli operators \cite{nielsenandchuang}. The controlled-not, {\it CNOT}, is an example of a two qubit gate.  {\it CNOT} performs an X gate on the second qubit only if the first qubit is in state $\ket{1}$. It is well-known that every single-qubit operation can be expressed in terms of the parameterized $U3$ gate.
{\footnotesize
$$U3(\theta, \phi, \lambda) = \begin{pmatrix} \cos(\theta/2) & -e^{i\lambda}\sin(\theta/2)\\
\sin(\theta/2) & e^{i\phi + i\lambda}\cos(\theta/2) \end{pmatrix}$$
}
\parah{Quantum Programs}
A quantum program can be expressed as a single operator on an arbitrary number of qubits, while hardware implements a very small set of single- and two-qubit\footnote{Superconducting qubits have two qubit gates currently, trapped ion qubits can implement small degree all-to-all gates.} gates. 

A quantum program is a finite sequence of unitary operators of the form $U^1_{Q_1}U^2_{Q_2}\ldots U^d_{Q_d}$ applied to
a system of qubits. Here $U_{Q_i}$ is a unitary operator applied to the subset of qubits $Q_i$. For example, a quantum program that prepares a bell state, $\frac{1}{\sqrt{2}}(\ket{00} + \ket{11})$, is given by $\{H_{\{0\}}, CNOT_{\{0,1\}}\}$. Graphically, we represent quantum programs as circuits where the wires represent qubits evolving through time from left to right. See Figure \ref{fig:bellstate} for an example.

A quantum program is an operator, so it can be represented as a unitary matrix.  The unitary representation of a quantum program written as a sequence of gates can be obtained as follows:
First, all gates are lifted to the number of qubits involved in the program.  For example, the single-qubit gate $H_{\{0\}}$ in Figure \ref{fig:bellstate} can be lifted to a two-qubit gate by taking the tensor product of the gate with the identity gate that is denoted by the $2\times2$ identity matrix $I_2$.  That is the lifted two-qubit gate is $H_{\{0\}} \otimes I_2$.
The order here implies that the Hadamard gate (i.e. $H_{\{0\}}$), is applied to the first qubit and the identity or no-op is applied to the second qubit.
Once all gates in the program have been lifted to the same dimension, the product of the lifted matrices yields the unitary representation of the program.  We will use the notation ${\rm\it Compose}(U^1_{Q_1}U^2_{Q_2}\ldots U^d_{Q_d})$ to denote the unitary matrix for the program $U^1_{Q_1}U^2_{Q_2}\ldots U^d_{Q_d}$.
{\footnotesize
\begin{figure}
    \centering
    \includegraphics[scale=0.7]{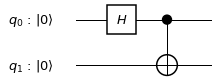}
    \caption{\footnotesize \it A circuit diagram for a Bell State Preparation program.
    The qubits, $q_0$ and $q_1$, are both prepared in the $\ket{0}$ state.
    A Hadamard operation is applied to $q_0$ resulting in $q_0$ being in the $\ket{+} = \frac{1}{\sqrt{2}}(\ket{0} + \ket{1})$ state.
    This is followed by a controlled-not operation from $q_0$ to $q_1$. The final state is $\frac{1}{\sqrt{2}}(\ket{00} + \ket{11})$ often referred to as a Bell state.}
    \label{fig:bellstate}
\end{figure}
}

\parah{Distinguishability} Distinguishability is centered on determining closeness
for quantum states or operators. State fidelity is a measure of similarity between two quantum states.  It will return a probability that one state can pass a test to identity as the other.
Given two quantum pure states, $\ket{\rho}$ and $\ket{\psi}$, their state fidelity
is defined by $|\braket{\rho\mid\psi}|^2$, where $\braket{\rho\mid\psi}$ is the standard inner product between $\ket{\rho}$ and $\ket{\psi}$. A fidelity of 1 corresponds to equal states, where as a fidelity of 0 corresponds to opposite states.

For distinguishability between quantum operators (or quantum programs or gates), a measure of unitary distance is used. Recently, most synthesis tools have been using the Hilbert-Schmidt inner product to compute closeness~\cite{davis2019heuristics,qaqc}. Given two unitary operations $U_1$ and $U_2$, the Hilbert-Schmidt inner product is defined as $\braket{U_1, U_2} = {\rm\it Tr}(U_1^{\dagger}U_2)$. ${\rm\it Tr}$ here is the matrix trace function which is defined as $Tr(X) = \sum_i^d{X_{ii}}$, where $X$ is a $d\times d$ matrix.  


\parah{Synthesis} Given a quantum program as a unitary matrix $U$, how can we come up with a quantum program $U^1_{Q_1}U^2_{Q_2}\ldots U^d_{Q_d}$ such that:
\begin{enumerate}
\item each $U^i_{Q_i}$ is a quantum gate,
\item ${\rm\it Compose}(U^1_{Q_1}U^2_{Q_2}\ldots U^d_{Q_d}) = Q$, and
\item $d$ is minimal.
\end{enumerate}

The $U^i_{Q_i}$s are picked from a fixed and finite set of gates.  The set of gates are determined by the underlying hardware. Furthermore, effective synthesis tools produce short circuits. This is because longer circuits accumulate more noise resulting in a larger error in the final output. 

\parah{Topology} Performing quantum operations on hardware can involve more compilation steps than synthesis. After synthesis, a target quantum operation has been broken down into a sequence of gates. However, not every two-qubit operation can be directly executed on the hardware. The device's coupling map or topology defines a graph of qubit interactions, see Figure \ref{fig:topo} for an example. The nodes in this graph represent physical or device qubits, and the edges represent possible interactions. If a quantum operation requires a two-qubit gate between two qubits not connected by an edge, routing operations will need to be inserted to perform the gate. This process is called mapping and has significant overhead on circuit depth. However, a synthesis algorithm can be topology-aware: all gates produced are directly executable on the device without need for extra routing operations.

%% file: sections/circuit.tex
\section{Continuous Representation of a Circuit}
\label{sec:circuit}

QFAST relies heavily on the encoding that captures the application of a unitary to an arbitrary subset of qubits within a circuit. 
This uses Pauli matrices and the Lie Group Structure of $U(n)$, the group of $n\times n$ unitary matrices.

\subsubsection{Lie Group Structure of $U(n)$.}
The group of $U(2)$ is the Lie group of unitary $2\times 2$ matrices.
It's Lie algebra $\mathfrak{u}(2)$ is the set of $2\times 2$ skew-Hermitian matrices.
The Lie algebra $\mathfrak{u}(2)$ is spanned by the set $\{\I\sigma_i, \I\sigma_x, \I\sigma_y, \I\sigma_z\}$, where $\I$ is the imaginary number, $\sigma_i$ is the $2\times 2$ identity matrix, and $\sigma_x, \sigma_y, \sigma_z$ are the Pauli matrices $X, Y, Z$, respectively from Section \ref{background}.  
The infinitesimal generators of $U(2)$ can be given by the set $\{\I\sigma_i, \I\sigma_x, \I\sigma_y, \I\sigma_z\}$.
We are interested in these generators because any one-qubit operator can be written as the matrix exponential of a linear combination of the generators:
$$U(2) = \{e^{\I(\vec{\alpha} \cdot \vec{\sigma})} \mid\ \vec{\alpha} \in \mathbb{R}^4\}$$
\noindent In other words, each one-qubit gate can generated by picking a suitable value for $\vec{\alpha}$.  Alternatively, one can see $U(2)$ as a parametric representation of any single-qubit gate.

In the following discussion, let $\vec{\sigma} = \{\sigma_i, \sigma_x, \sigma_y, \sigma_z\}$.  The construction of $U(2)$ generalizes to the group of $2^n\times 2^n$ unitary matrices $U(2^n)$ as follows.  The Lie algebra $\mathfrak{u}(2^n)$ is the set of $2^n\times 2^n$ skew-Hermitian matrices. Similarly, we can generate all $2^n\times 2^n$ Hermitian matrices with the nth-order Pauli matrices:
$$\vec{\sigma}^{\otimes n} = \{\sigma_j \otimes \sigma_k \mid \sigma_j \in \vec{\sigma}, \sigma_k \in \vec{\sigma}^{\otimes n-1} \}$$
Consequently, we get a similar construction of $U(2^n)$:
$$U(2^n) = \{e^{\I(\vec{\alpha} \cdot \vec{\sigma^{\otimes n}})} \mid\ \vec{\alpha} \in \mathbb{R}^{4^n}\}$$

$U(2^n)$ provides us a continuous representation of all quantum operators on $n$-qubits. While there are many ways to represent the unitary group, we choose this representation because of its operational meaning. We can characterize a quantum operator by its corresponding element in the Lie algebra $\mathfrak{u}(2^n)$ decomposed in the Pauli basis $\sigma^{\otimes n}$.  

In order to produce a continuous representation of a circuit, we need to be able to structure gates that are only applied to a subset of qubits. We can use this idea to quickly produce n-qubit operators that only act on a subset of the n-qubits. We simply restrict the elements of the nth-order Pauli basis to those elements that have $\sigma_i$ in all the positions where those qubits should be left untouched. For example, a two-qubit quantum operator generated only by $\sigma_x \otimes \sigma_i$ acts only on the first qubit. 
Furthermore, this operator can be rewritten as a single-qubit operator generated by $\sigma_x$ with the same coefficient:
$$e^{ i ( \alpha_x * (\sigma_x \otimes \sigma_i) ) } = e^{ i (\alpha_x * \sigma_x) } \otimes \sigma_i$$

For another example, suppose we want to produce a general 4-qubit gate that is applied only to qubits 0 and 2. To accomplish this, we restrict the 4th-order Pauli basis to those which have $\sigma_i$ at positions 1 and 3:
{\footnotesize
\[
\{\sigma_{iiii}, \sigma_{iixi}, \sigma_{iiyi}, \sigma_{iizi}, \sigma_{xiii}, \sigma_{xixi}, \sigma_{xiyi}, \sigma_{xizi}, \sigma_{yiii}, \sigma_{yixi}, \sigma_{yiyi}, \sigma_{yizi}, \sigma_{ziii}, \sigma_{zixi}, \sigma_{ziyi}, \sigma_{zizi}\},
\]}where we use $\sigma_{jklm}$ to denote $\sigma_j \otimes \sigma_k \otimes \sigma_l \otimes \sigma_m$

These 16 Pauli matrices are a subset of the 4th-order Pauli matrices. Any operator that is produced by exponentiating a real linear combination of these, after multiplying $\I$ throughout, will only affect qubits 0 and 2. There are 16 real parameters. The operator produced is a $16\times 16$ unitary since this was constructed from 4th-order Pauli's.
However, we can quickly extract the 2-qubit operator by copying the coefficients similar to the previous example.
With this in mind, we can construct a general n-qubit gate that acts only on m-qubits where $m \leq n$. If we fix the qubits we wish to operate on, this produces a continuous construction of a gate on these qubits.


$$l_0(\alpha_0\sigma_{ii} + \alpha_1\sigma_{xi} + \alpha_2\sigma_{yi} + \alpha_3\sigma_{zi}) + l_1(\alpha_0\sigma_{ii} + \alpha_1\sigma_{ix} + \alpha_2\sigma_{iy} + \alpha_3\sigma_{iz})$$

We can now generalize this construct to gates of size $m$ in an $n$-qubit system, with $m \le n$. To start, we show how we can restrict the $n$-th order Pauli basis. We define the set $P_l^{\otimes n}$, which contains all the $n^{\rm th}$ order Pauli's with the identity matrix in the $l^{\rm th}$ position in tensor order:

$$P_l^{\otimes n} = \{ \sigma_j \otimes \sigma_i \otimes \sigma_k \mid \sigma_j \in \vec{\sigma}^{\otimes l}, \sigma_k \in \vec{\sigma}^{\otimes (n - 1 - l)} \}$$

Using this we can restrict the Pauli basis by a set of qubits $Q$:

$$\vec{\sigma}^{\otimes n}_Q = \{ \sigma_k \mid \sigma_k \in \vec{\sigma}^{\otimes n} \mbox{ and } \forall j \notin Q:\ \sigma_k \in P_j^{\otimes n}\}$$

The above notation is parametric with respect to a set of qubits $Q$.  How can we generalize the construct to any subset of qubits where the cardinality of each subset is $m$?  For this we introduce a vector of indicator variables $\vec{l}$.  Exactly one element in the vector should be 1 and the rest should be 0.  If an element, say $l_Q$, of $\vec{l}$ is $1$, then we get a parametric operator that is applied to the qubits in $Q$.  
We can define a continuous, generic gate in terms of all subsets of m qubits from an n-qubit system using the indicator variables as follows:

$$G^{\otimes n}_m(\vec{\alpha}, \vec{l}) = \mbox{\Large $e$}^{\I\sum_{|Q| = m}{\frac{e^{l_Q}}{\sum_i{e^{l_i}}}(\vec{\alpha} \cdot \vec{\sigma}^{\otimes n}_Q)}}$$

Note that the outer sum ranges over all subsets of $m$-qubits. $G^{\otimes n}_m(\vec{\alpha}, \vec{l})$ is a $2^n\times 2^n$ unitary matrix that represents a generic quantum operator affecting only $m$-qubits.  We apply exponent to each element of $\vec{l}$ so that the space of values assumed by each element is continuous. $G^{\otimes n}_m(\vec{\alpha}, \vec{l})$ is parametric with respect to $\vec{\alpha}, \vec{l}$.  An  assignment to $\vec{\alpha}, \vec{l}$ gives a single instance of a gate operating on a set of $m$ qubits.  

The generic representation of an arbitrary gate can be generalized to a circuit as follows. All $n$-qubit circuits composed of $d$ $m$-qubit gates can be described by the product of the generic gates:

$$\prod_{i=1}^d{G^{\otimes n}_m(\vec{\alpha}^{(i)},\vec{l}^{(i)})}$$

Finally, we introduce another notation which fix the location of a generic gate by choosing the active qubits $Q$ and removing the other terms:

$$F^{\otimes n}_m(\vec{\alpha}, Q) = \mbox{\Large $e$}^{\I(\vec{\alpha} \cdot \vec{\sigma}^{\otimes n}_Q)}$$

%% file: sections/algorithm2.tex
\section{QFAST Hierarchical Synthesis}\label{sec:alg}



We propose a hierarchical approach to synthesis that uses iterative refinement. 
As low depth is of importance and best published methods~\cite{kak,ionsynth,davis2019heuristics,qaqc} use numerical optimization we have decided apriori for this formulation.
These techniques build up a circuit layer-by-layer~\cite{kak,ionsynth,davis2019heuristics,qaqc}. At each step, a layer is added using two-qubit building blocks composed of single- and two- qubit native gates: single qubit gates are parameterized (e.g. generic U3 gate), but two-qubit gates are non-parameterized functions (e.g. {\em CNOT}). When a layer is added, multiple placements for a single block are evaluated. The process continues to the net effect of building a tree of partial solutions, where each node is a partial solution, each edge is the placement of an additional building block and each node is evaluated individually.

The algorithms differ in the structure of the basic building block and the strategy to expand the partial solution tree. However, since the basic building blocks are limited in the function they can perform and multiple placements need to be evaluated, these algorithms seem to be slow due to the combinatorial number of evaluated partial solutions.  Exploiting parallelism  in walking the tree has been explored as a solution to improve execution time, but a more intrinsic scalability challenge may still remain. As any partial solution can be the final solution, a very stringent numerical optimization is employed at each step: the constraint is that each partial solution has to be numerically optimized with a minuscule  distance from target. Rephrased in Quantum Information Science (QIS) terminology, at each step they attempt to make the partial solution indistinguishable from the target. This is compounded by the fact that the search may descend very deep in the tree before backtracking or moving laterally in a ``breadth-first'' direction. Deep partial solutions have a large number of parameters and work is wasted if backtracking or ``lateral'' (breadth-first) moves occur.

QFAST tries to address these shortcomings through very  simple intuitive principles:
\begin{enumerate}
\item  As small two-qubit building blocks may lack ``computational power'', {\it we use generic blocks spanning a configurable number of qubits}.
\item As the number of partial solutions and their evaluation may hamper scalability, {\it we conflate the numerical optimization and frontier expansion}. At each step, the circuit is expanded by one layer. Given a $n$ qubit circuit, a layer encodes an ``arbitrary'' operation on any $m$ qubits, with $m < n$.
  Thus, our formulation solves only $O(d)$ optimization problems, where $d$ is the solution depth. Note that during this process, once a block is placed at a certain depth, the algorithm has the liberty of choosing and reassigning the subset of qubits it operates on. We refer to this stage as {\it expansion}. 
\item As numerical optimization speed is proportional with the ``quality'' of the solution, {\it we built the algorithm to solve less constrained problems}. This translates into having most  of each expansion step  look for a ``large'' value for the distance to solution. This computes an approximation of  the structure and the depth of circuit  that  gets close
  enough to the solution. This results in easier and faster-to-solve problems for optimizers. Once structure is fixed, we then refine the ``function'' and attempt optimization with a stringent distance. 
  \end{enumerate}

{\footnotesize
\begin{figure}
    \centering
    \includegraphics[width=\textwidth]{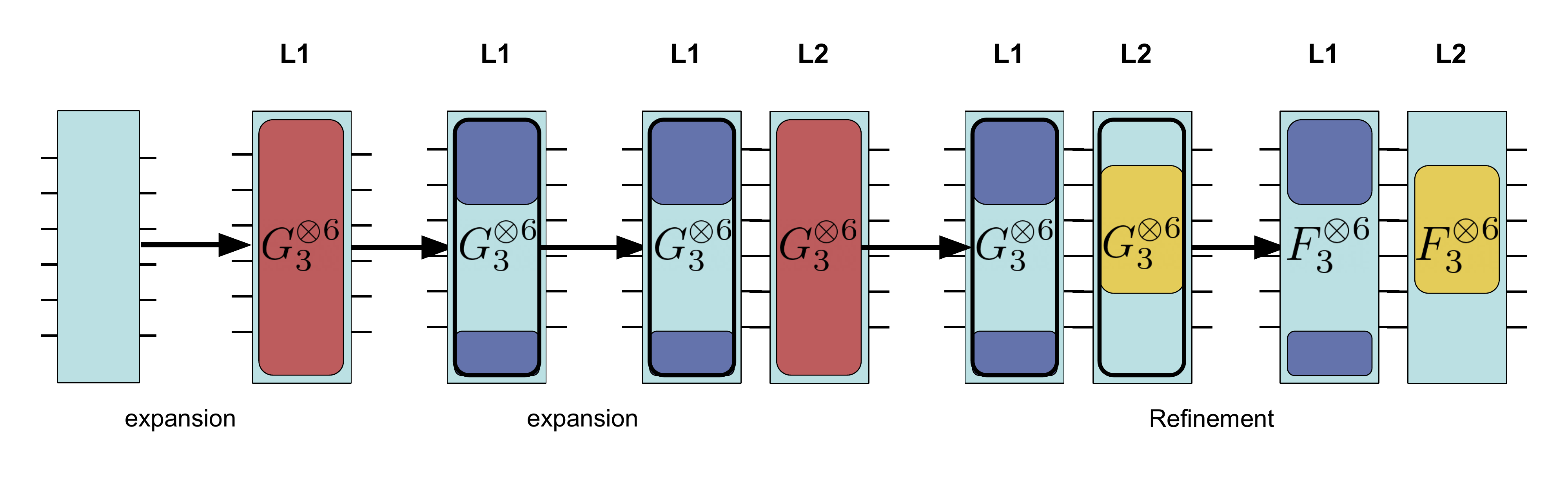}
    \caption{\label{fig:steps} \footnotesize \it An example walk though of QFAST's exploration stage. Initially, the empty circuit is expanded to the first generic gate. Here there are 6 qubits, and the target block size is 3. The first invocation of an optimization initializes parameters for the gate, and then the second expansion occurs. Again the optimizer is invoked on the entire circuit initializing the variables. Once a candidate solution is found, refinement fixes the location of the generics, producing F-type gates, and reducing the final solution distance.}
\end{figure}
}






\subsection{QFAST algorithm}

Starting with a $n$ qubit unitary, the algorithm breaks down a unitary into a product of smaller unitaries in a hierarchical manner.
It starts by solving for a circuit in terms of $\frac{n}{2}$-qubit operators, $G^{\otimes n}_{\frac{n}{2}}$. Then it expands each $\frac{n}{2}$-qubit operator into $\frac{n}{4}$-qubit operators, and so on. During this {\it expansion} process, we maintain the association between blocks and qubits.
Expansion produces circuits composed of generic building blocks. At some point, the algorithm has to switch into a mode where these blocks are further specialized  using single- and two-qubit gates native to the quantum processor. This stage is referred to as {\tt instantiation}. 
In  {\tt instantiation}, all the generated ``small'' blocks are transformed into circuits composed of native gates directly executable on the quantum processor. The final stage, {\tt recombination} stitches all the executable blocks, walking back the hierarchy generated during {\tt expansion} and places the native gates on right qubits at the right time sequence.

\parah{Expansion} Expansion grows the circuit layer by layer, until its distance is close enough to the target input unitary. This expansion phases works by first exploring circuit structure and then, once a candidate solution is found, refining the result. {\tt Exploration} is responsible for the growing of the circuit. This determines an initial structure and function. Each invocation of {\tt exploration} starts with the result from the previous invocation with an additional unbound operator $G^{\otimes n}_m$ and tries to solve for all variables.
This is how we conflate search for structure and function\footnote{In this case, structure means the application of gates to qubits, rather than the values of our parameters.}  with numerical optimization. {\tt Refinement} is responsible for reducing the distance to a final acceptable level.

\parah{Exploration} In exploration we instantiate an optimizer with a large learning rate. This serves an important purpose. At this stage both structure and function are undetermined and we need to solve an optimization problem with a large number of parameters. A fast moving optimizer will quickly search over many possible configurations. Furthermore, having a coarse success criteria reduces execution time. A candidate solution can then be sent to refinement to be made acceptable. 
The target criteria or distance {\tt $exploration\_distance$} is a customizable parameter that is optimizer specific.

During exploration, optimizer progress is of concern and we need to preclude performing a large number of iterations that do not improve the quality of the solution.
Every 20 optimizer steps we record the value of the loss function. If the last 100 recorded loss values haven't changed much, we determine that we have plateaued and stop the optimizer. We record the values of all variables, add another layer of gates to the circuit and reinstantiate the variables we have seen with the values recorded. This process is done in a loop until we observe a loss value below the {\tt $exploration\_distance$}. At this point we refine the circuit. 

\parah{Refinement Stage}
Exploration produces a sequence of $G^{\otimes n}_m$'s and produces the numerical value of all parameters. Their solutions are numerically instantiated for both function and structure. On the other hand, due to the coarse criteria, the function is just a coarse approximation of the target computation. Thus, we need to further refine our result to provide a more acceptable error/distance value. To accomplish this, we use the $F^{\otimes n}_m$ encoding of the circuit. The structure parameters are seeded and fixed using the exploration numerical result. We then pass the circuit back into the optimizer with a much smaller learning rate. The optimizer is now enabled to refine the solution down to a much lower distance, denoted by ${\tt refinement\_distance}$. 

\parah{Instantiation} The expansion stage produces a candidate circuit composed of generic blocks.
While these can perform any computation, they are not directly executable on hardware. Thus, we need a stage where blocks are transformed and rewritten into hardware native gates. At this stage, we can leverage previous approaches. KAK~\cite{kak} decomposition is an ubiquitous technique deployed in commercial compilers, and it generates depth optimal circuits for two qubit unitaries. Thus, after exploration reaches the two qubit level, QFAST applies KAK on all blocks. 
Furthermore, the hierarchical nature of QFAST gives us an opportunity to compose with other synthesis algorithms at any granularity.
For example, we have QFAST instantiations that apply UniversalQ~\cite{uq} at arbitrary levels. 

\subsection{Loss Function and Solution Distance}

The goal of synthesis is to find $U_C$ such that it minimizes $\Delta(U_C, U_T)$, where the $U_C$ is the operation implemented by the encoded circuit, $U_T$ is the target input, and $\Delta$ is some unitary distance function. Ideally, we find $\Delta(U_C, U_T) = 0$. However, due to numerical floating point arithmetic constraints and optimizer limitations we  attempt to  find $U_C$ that satisfies $\Delta(U_C, U_T) < \epsilon$ for some acceptable threshold $\epsilon$. 

We use the Hilbert-Schmidt inner product in our distance function:

$$\langle U_C, U_T \rangle = Tr(U_C^{\dagger}U_T)$$

The closer that $U_C$ and $U_T$ become, the closer  the product $U_C^{\dagger}U_T$ is to the identity matrix. As the product approaches identity, its trace  becomes closer to the dimension, $d$, of the matrix. Our completed distance function normalizes the value of the inner product to be within the range of $(0, 1)$. Lastly, we invert it, so that a value of 0 correspondes to an exact match and a value of 1 implies the opposite. This allows us to treat $\Delta$ as a loss function and invoke an optimizer's minimize routine on it. The final function is given by:

$$\Delta(U_C, U_T) = \sqrt{1 - \frac{|Tr(U_C^{\dagger}U_T)|^2}{d^2}}$$

During the exploration stage we use a fast optimization scheme designed to quickly find the circuit structure. The ${\tt exploration\_distance}$ threshold for this stage is optimizer specific and it has been determined empirically to provide a good combination of speed and quality of solution. As optimizers are very unpredictable, there is probably no procedure to determine this value from first principles. Our default setting is ${\tt exploration\_distance} = 0.02$. We note that probably contrary to intuition, lowering this value results in longer circuits. The optimizer reaches the solution but it requires more iterations to compute the parameters. Since we try to detect and avoid plateaus, we give preference  to adding  another layer instead of slow convergence.

The refinement step fixes circuit structure and uses a better but slower optimizer to reduce the error down as low as possible, stopping if it falls below the ${\tt refinement\_threshold}$. Again, this value needs to be determined empirically and in our experiments we use a stopping criteria  ${\tt refinement\_threshold} = 10^{-5}$. As indicated by the results, the final value is in practice much lower, which indicates that  tighter  values are possible.

%% file: sections/props.tex





\subsection{Topology Awareness}

The QFAST formulation allows for topology-aware synthesis. As shown by
Davis et al~\cite{davis2019heuristics}, this is required to obtain short
circuits as third party compilers, optimizers, and mappers cannot offset the
loss of quality when topology awareness is missing.


Topology is easily incorporated into QFAST using the continuous
gate/circuit representation, which encodes structure, i.e. the qubits
the gate operates on. Assuming all-to-all connectivity, the
the $G^{\otimes n}_m$ representation will have $\binom{n}{m}$ parameters to
encode all possible placements. For restricted connectivity all we
have to do is generate only the terms that correspond to all strongly
connected components of size $m$ in the $n$ target device's coupling graph.
Figure~\ref{fig:topo} illustrates this for an example where a four
qubit gate ($n = 4$) is expanded into two qubit ($m = 2$) blocks. With all-to-all
connectivity we will have to generate six $l_i$ variables, while after
pruning for topology we generate only three, corresponding to the links
$(q_0, q_1), (q_1, q_2)$ and $(q_1, q_3)$. 

{\footnotesize
\begin{figure}
    \centering
    \includegraphics[scale=0.4]{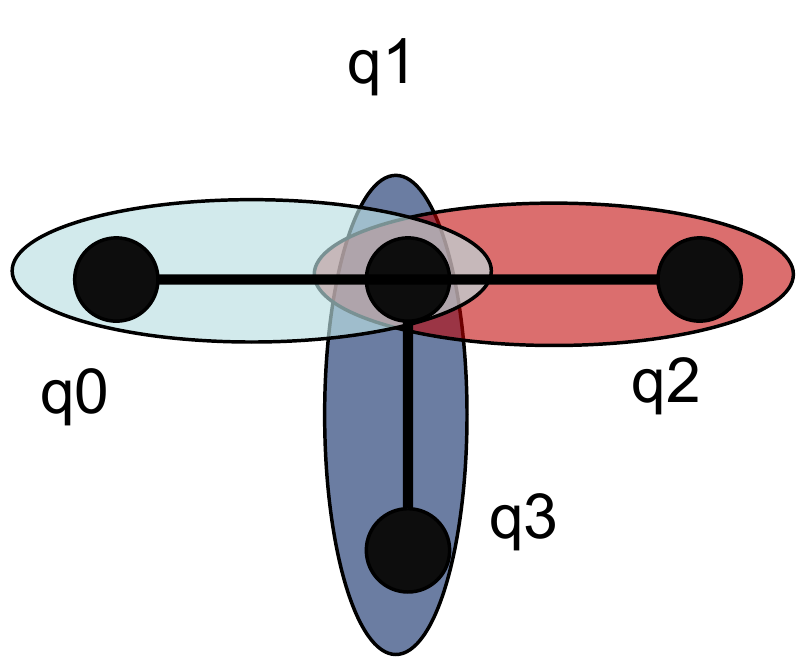}
    \caption{\label{fig:topo} \footnotesize \it An Example 4-qubit topology. We can make QFAST topology-aware by restricting the possible placements to all strongly connected components in the topology.}
\end{figure}
}

\subsection{Complexity Analysis}
\label{sec:compl}

For a $n$ qubit target unitary, $m$ qubit block size, with $m < n$, and
an all-to-all topology, the space complexity of our variable-location
generic gate encoding is given by $O(\binom{n}{m} + 4^{m})$.
While for other topologies, $\mathcal{T}$, we'll have
$O(SCC(\mathcal{T}, m) + 4^{m})$ space complexity,
where $SCC(\mathcal{T}, m)$ denotes the number of strongly connected
components of size m within the larger n-graph topology.
When we fix structure, our space complexity shrinks to $O(4^{m})$.
Finally, the space complexity of a circuit of depth $d$ simply adds $d$ as a factor.



%% file: sections/eval.tex
\section{Evaluation}
\label{sec:eval}

\subsection{Software Implementation}

We implemented QFAST in Python 3.6 using TensorFlow 1.13.1 for
encoding the circuit structure and loss function. We call the
ADAMOptimizer package from TensorFlow to minimize the loss
function. QFAST experiments ran on a single node of the
Cori supercomputer hosted at the National Energy Research
Scientific Computing Center (NERSC), where nodes contain
two Intel Xeon E5-2698 v3 ("Haswell") processors at 2.3 GHz
(32 cores total).  Both software and benchmarks are open source,
location witheld for blind review purposes.
We use the IBM QISkit software to perform the
KAK 2-qubit decomposition during the
instantiation stage of QFAST.

\subsection{Benchmarks}

Our benchmark suite contains small to medium circuits and algorithms
appropriate for the NISQ era, used previously by other
researchers~\cite{cowtan2019phase,davis2019heuristics,murali2019noise}.
There are several classes of circuits. First are optimal depth, some
taken from literature (Peres, Fredkin, multi-qubit control gates),
some generated by specialized domain generators~\cite{qiskitaqua} (Grover, QFT).

The second class include Variational Quantum Eigensolver
(VQE)~\cite{McClean2015} circuits generated for chemistry by
OpenFermion~\cite{openfermion}. VQE is currently perceived as one of the most
promising algorithms to deliever on the transformational promise of
quantum computing. VQE circuits are parameterized and the algorithm
variationally updates the parameters. The circuit executes, the result
is passed into a classical optimizer which recomputes the circuit
parameterization, the circuit is updated and the cycle continues until
the chemistry solution is found. VQE circuits are fixed depth and
there are no first principle approaches (domain generators) to
specialize for the intermmediate results/circuits.

The third class of circuits are generated for problems that study the
time evolution of chemical systems, such as Transverse Field Ising
Model (TFIM)~\cite{tfimshin,tfimlb}. TFIM is an exponent of chemical simulations
using time dependent Hamiltonians. In this case, domain generators append a
fixed function block per step and  circuit depth grows
linearily. Domain generators concentrate in reducing ``block'' depth
and can't avoid linear growth.

On all classes of circuits, traditional compilers
fail~\cite{McClean2015,tfimlb} to reduce circuit depth. The apriori optimal
circuits are a worst case test scenario for synthesis, as it can only
match or increase depth.  The other two, one fixed depth, other ever
increasing are good candidates to showcase the value of synthesis
tools.

\subsection{Evaluation Criteria}



The criteria we are most interested in is the depth of the generated
circuit. To place QFAST in context, we evaluate against the compiler
presented by Davis et al~\cite{davis2019heuristics}, referred to as the {\it
  SearchCompiler}.  {\it SearchCompiler} claims to produces depth optimal
circuits, but execution does not seem to scale above four qubits. We
also evaluate against UniversalQ~\cite{uq} (UQ), the state-of-the-art
compiler based on linear algebra approaches. {\it SearchCompiler} is
topology-aware, while for UniversalQ topology seems to increase the
circuit depth. 

To even the comparison, we assume in all experiments all-to-all
chip connectivity. As discussed in Section~\ref{sec:compl} this is the
worst case running time for QFAST.  It is also the best case for UQ in
depth and performance.

While interested in the running time of QFAST, we note that none of
its implementation is tuned for performance. We execute on Intel CPUs,
while TensorFlow can run much faster on GPUs. Furthermore, we did not
attempt to 
exploit distributed memory parallelism in TensorFlow.

Results are summarized in Figures \ref{tab:3qub}, \ref{tab:4qub}, \ref{tab:56qub}.

\input{sections/table.tex}

\subsection{Circuit Depth and Solution Quality}

When applied to circuits where an optimal depth implementation is
known, QFAST is clearly sub-optimal and increases depth on average by
$4.3 \times$ and up to
$10\times$. {\it SearchCompiler} matched the optimal depth for most three
qubit circuits, but we could not obtain any results for any of the
four or greater qubit benchmarks due to numerical errors or timeouts after 24
hours of execution. When applied on optimal circuits UniversalQ
increases depth on average by $12 \times$ and up to $60 \times$. 

When applied to VQE and TFIM circuits QFAST improves depth on average
by $6.3 \times$ and up to $30 \times$. On the same circuits, UniversalQ
improves depth on average by $1.5 \times$ and up to $6.6 \times$. For any
circuit of four qubits or more, QFAST generated shorter solutions than UniversalQ.

Tables~\ref{tab:3qub},~\ref{tab:4qub},~and~\ref{tab:56qub} shows that QFAST produces circuits at a distance
from the target unitary ranging from $10^{-3}$ to $10^{-7}$, {\it
  SearchCompiler} roughly at $10^{-7}$ and UniversalQ roughly at
$10^{-8}$. To test the quality of the circuits we have run simulations
with inputs set to all
the standard basis state vectors and 1000 random state vectors.  For
all circuits with a distance less than $10^{-3}$, the average output state fidelity
is in the range 0.9999..., with ULP difference of $10^{-5}$
digit. UniversalQ fidelities are in the range 0.9999999999999..., with ULP
$10^{-13}$ difference of digit.

%% file: sections/table.tex
{\footnotesize\tiny
  \begin{figure}
    \centering
    \makebox[\textwidth][c]{\begin{tabular}{|c|c|c|c|c|c|c|c|c|c|c|c|c|c|c|c|c|c|}
\hline
\multicolumn{3}{|c|}{ Benchmark } & \multicolumn{3}{c|}{ QFAST + KAK } & \multicolumn{3}{c|}{ UniversalQ } & \multicolumn{3}{c|}{ Search Compiler }\\
\hline
Name & n & Depth & Depth & Distance & Time (s) & Depth & Distance & Time (s) & Depth & Distance & Time (s)\\
\hline
ccx & 3 & 6 & 42 & $1.4 \times 10^{-6}$ & 1395.1 & 15 & $2.6 \times 10^{-8}$ & 0.2 & 8 & $2.4 \times 10^{-7}$ & 576.1 \\
\hline
fredkin & 3 & 8 & 33 & $2.2 \times 10^{-6}$ & 1163.5 & 14 & $0$ & 0.2 & 8 & $5.8 \times 10^{-6}$ & 433.8 \\
\hline
grover\_s01 & 3 & 7 & 14 & $8.1 \times 10^{-7}$ & 97.6 & 20 & $0$ & 0.2 & 7 & $5.5 \times 10^{-7}$ & 315.5 \\
\hline
or & 3 & 6 & 15 & $6.5 \times 10^{-7}$ & 171.3 & 15 & $2.6 \times 10^{-8}$ & 0.2 & 8 & $5.8 \times 10^{-7}$ & 587.9 \\
\hline
peres & 3 & 5 & 18 & $6.8 \times 10^{-7}$ & 688.1 & 13 & $2.1 \times 10^{-8}$ & 0.2 & 7 & $2.3 \times 10^{-7}$ & 309.6 \\
\hline
qft3 & 3 & 6 & 6 & $3.0 \times 10^{-7}$ & 50.0 & 15 & $3.0 \times 10^{-8}$ & 0.2 & 6 & $4.9 \times 10^{-7}$ & 202.5 \\
\hline
\end{tabular}}
\caption{\label{tab:3qub} Summary of results for 3-qubit benchmarks. QFAST compiled the 3-qubit benchmarks down to blocks of 2-qubits and then instantiated with KAK. QFAST is compared against {\it Search Compiler} and {\it UniversalQ}. The depth columns denote the number of CNOTs in the circuit.}
\end{figure}
  \begin{figure}
    \centering
    \makebox[\textwidth][c]{\begin{tabular}{|c|c|c|c|c|c|c|c|c|c|c|c|c|c|c|c|c|c|}
\hline
\multicolumn{3}{|c|}{ Benchmark } & \multicolumn{3}{c|}{ QFAST + KAK } & \multicolumn{3}{c|}{ QFAST + UQ } & \multicolumn{3}{c|}{ UniversalQ }\\
\hline
Name & n & Depth & Depth & Distance & Time (s) & Depth & Distance & Time (s) & Depth & Distance & Time (s)\\
\hline
TFIM-1 & 4 & 6 & 8 & $6.0 \times 10^{-7}$ & 67.3 & 80 & 60.3 & $5.5 \times 10^{-7}$ & 82 & $2.1 \times 10^{-8}$ & 0.6 \\
\hline
TFIM-10 & 4 & 60 & 24 & $9.5 \times 10^{-4}$ & 1286.4 & 80 & 78.3 & $3.7 \times 10^{-3}$ & 95 & $3.0 \times 10^{-8}$ & 0.6 \\
\hline
TFIM-22 & 4 & 126 & 21 & $1.0 \times 10^{-5}$ & 1187.5 & 100 & 231.1 & $5.4 \times 10^{-3}$ & 85 & $4.2 \times 10^{-8}$ & 0.7 \\
\hline
TFIM-35 & 4 & 210 & 16 & $8.8 \times 10^{-7}$ & 225.4 & 80 & 461.2 & $3.4 \times 10^{-6}$ & 97 & $4.2 \times 10^{-8}$ & 0.6 \\
\hline
TFIM-60 & 4 & 360 & 55 & $1.5 \times 10^{-6}$ & 1529.7 & 80 & 148.6 & $6.2 \times 10^{-7}$ & 93 & $2.6 \times 10^{-8}$ & 0.6 \\
\hline
TFIM-80 & 4 & 480 & 40 & $1.7 \times 10^{-6}$ & 1248.4 & 80 & 126.0 & $6.9 \times 10^{-7}$ & 89 & $2.1 \times 10^{-8}$ & 0.6 \\
\hline
TFIM-95 & 4 & 570 & 17 & $7.1 \times 10^{-7}$ & 280.2 & 80 & 169.4 & $9.3 \times 10^{-7}$ & 91 & $2.1 \times 10^{-8}$ & 0.7 \\
\hline
TFIM-100 & 4 & 600 & 17 & $9.2 \times 10^{-7}$ & 277.4 & 80 & 142.3 & $9.9 \times 10^{-7}$ & 91 & $6.1 \times 10^{-8}$ & 0.6 \\
\hline
Ethy-1 & 4 & 64 & 37 & $8.8 \times 10^{-6}$ & 1226.2 & 100 & 1473.7 & $1.0 \times 10^{-6}$ & 99 & $4.7 \times 10^{-8}$ & 0.6 \\
\hline
Ethy-2 & 4 & 64 & 30 & $4.5 \times 10^{-3}$ & 2192.7 & 39 & 225.6 & $3.6 \times 10^{-3}$ & 97 & $2.7 \times 10^{-8}$ & 0.6 \\
\hline
H2-1 & 4 & 56 & 5 & $7.2 \times 10^{-3}$ & 42.9 & 20 & 38.7 & $7.3 \times 10^{-3}$ & 92 & $2.1 \times 10^{-8}$ & 0.6 \\
\hline
H2-2 & 4 & 56 & 39 & $1.5 \times 10^{-3}$ & 2280.3 & 80 & 963.5 & $9.3 \times 10^{-3}$ & 98 & $4.2 \times 10^{-8}$ & 0.6 \\
\hline
qft4 & 4 & 12 & 21 & $7.9 \times 10^{-7}$ & 385.9 & 80 & 108.4 & $8.5 \times 10^{-7}$ & 85 & $3.9 \times 10^{-8}$ & 0.6 \\
\hline
bv & 4 & 3 & 18 & $5.8 \times 10^{-7}$ & 287.4 & 60 & 111.9 & $7.1 \times 10^{-7}$ & 91 & $3.0 \times 10^{-8}$ & 0.6 \\
\hline
cccx & 4 & 20 & 47 & $2.2 \times 10^{-5}$ & 2138.5 & 120 & 562.0 & $1.3 \times 10^{-6}$ & 70 & $2.1 \times 10^{-8}$ & 0.6 \\
\hline
\end{tabular}}
\caption{\label{tab:4qub} Summary of results for 4-qubit benchmarks. QFAST compiled the 4-qubit benchmarks down to blocks of 2-qubits and then instantiated with KAK. Additionally, QFAST compiled the 4-qubit benchmarks to blocks of 3-qubits and then instantiated with UQ. The depth columns denote the number of CNOTs in the circuit.}
\end{figure}

\begin{figure}
    \centering
\makebox[\textwidth][c]{
\begin{tabular}{|c|c|c|c|c|c|c|c|c|c|c|c|c|c|c|}
\hline
\multicolumn{3}{|c|}{ Benchmark } & \multicolumn{3}{c|}{ QFAST + UQ } & \multicolumn{3}{c|}{ UniversalQ }\\
\hline
Name & n & Depth & Depth & Distance & Time (s) & Depth & Distance & Time (s)\\
\hline
TFIM-10 & 5 & 80 & 120 & $1.2 \times 10^{-4}$ & 3994.2 & 429 & $3.0 \times 10^{-8}$ & 2.7 \\
\hline
TFIM-40 & 5 & 320 & 180 & $1.3 \times 10^{-6}$ & 1387.4 & 425 & $4.9 \times 10^{-8}$ & 2.7 \\
\hline
TFIM-60 & 5 & 480 & 180 & $1.5 \times 10^{-6}$ & 1409.8 & 425 & $7.7 \times 10^{-8}$ & 2.8 \\
\hline
TFIM-80 & 5 & 640 & 218 & $5.4 \times 10^{-5}$ & 3894.6 & 425 & $7.4 \times 10^{-8}$ & 2.7 \\
\hline
TFIM-100 & 5 & 800 & 280 & $1.6 \times 10^{-6}$ & 1264.9 & 429 & $4.2 \times 10^{-8}$ & 2.7 \\
\hline
TFIM-1 & 6 & 10 & 120 & $9.2 \times 10^{-7}$ & 1107.2 & 1794 & $3.7 \times 10^{-8}$ & 11.8 \\
\hline
TFIM-10 & 6 & 100 & 180 & $3.7 \times 10^{-3}$ & 7283 & 1809 & $8.7 \times 10^{-8}$ & 11.2 \\
\hline
TFIM-24 & 6 & 240 & 180 & $4.0 \times 10^{-3}$ & 7627.7 & 1803 & $7.6 \times 10^{-8}$ & 11.7 \\
\hline
TFIM-31 & 6 & 310 & 220 & $1.5 \times 10^{-3}$ & 12350.1 & 1797 & $4.9 \times 10^{-8}$ & 11.4 \\
\hline
TFIM-51 & 6 & 510 & 278 & $3.9 \times 10^{-3}$ & 10124 & 1819 & $5.2 \times 10^{-8}$ & 12.1 \\
\hline
Hubbard & 6 & 256 & 40 & $8.7 \times 10^{-4}$ & 532.8 & 1868 & $8.0 \times 10^{-8}$ & 12.6 \\
\hline
qft5 & 5 & 20 & 137 & $3.5 \times 10^{-6}$ & 5943.3 & 407 & $0$ & 2.7 \\
\hline
Grover\_s011 & 5 & 48 & 216 & $2.4 \times 10^{-6}$ & 3888.3 & 444 & $4.7 \times 10^{-8}$ & 2.7 \\
\hline
qft6 & 6 & 30 & 294 & $1.0 \times 10^{-6}$ & 19326 & 1777 & $5.6 \times 10^{-8}$ & 12.6 \\
\hline
\end{tabular}}
    \caption{\label{tab:56qub} Summary of results for 5-qubit and 6-qubit benchmarks. QFAST compiled the benchmarks down to blocks of 3-qubits and then instantiated with UQ. The depth columns denote the number of CNOTs in the circuit.}
    
  \end{figure}
  }

%% file: sections/discussion.tex
\section{Discussion} \label{disc}
Overall, we find the QFAST results encouraging for the future practical use of synthesis in quantum algorithm exploration in the NISQ era.
While not-optimal, we do improve upon previous synthesis techniques in either quality of solution or scalability.
The VQE and TFIM results show that QFAST can significantly reduce the depth of circuits used by domain scientists. These circuits are
the result of domain specific generators~\cite{openfermion,qiskitaqua,tfimlb} and QFAST can either displace efforts to optimize their functionality or provide much tighter bounds to guide their development.
Currently these circuits cannot be simplified by existing optimizing compilers, or by other synthesis packages.  The QFAST results indicate that synthesis on larger qubit blocks can be very useful inside the compiler optimization chain. Due to its composability and ability to use third party synthesis tools during instantiation we believe that QFAST is trivially portable to any new architecture and native gate set.

The data indicates that QFAST can generate shorter circuits provided the availability of third party optimal synthesis packages. We are communicating with the authors of {\it SearchCompiler} and expect a more robust release of their software. Upon availability, we expect to generate even shorter circuits and the results will motivate  further  development of optimal synthesis techniques specialized or scalable  up to a low number of qubits. 

We show scalability up to six qubits and as stated, we did not attempt to parallelize or accelerate the optimizer with GPUs. Without parallelization, scalability is limited by single node memory capacity. Our six qubit benchmarks ran on a server with 32 GB of memory, while a seven qubit benchmark ran out of memory on a server with 128 GB.  We know how to reduce the memory footprint of the algorithm and furthermore, parallelization will alleviate these constraints, as well as improve the execution speed. Since we are relying on the ADAMOptimizer package within TensorFlow we expect parallelization to be somewhat  painless.

%% file: related.tex
\section{Related Work}
\label{sec:related}

A foundational result  is provided by the
Solovay Kitaev (SK) theorem which 
relates circuit depth to the quality of the approximation~\cite{DawsonNielson05,Nagy16,ola15}. Different approaches~\cite{DawsonNielson05,ZXZ16,BocharovPRL12,MIM13,Qcompile16,ctmq,23gates,householderQ,CSD04,amy16,seroussi80}  have been
introduced since, with the goal of generating shorter depth circuits. 
These can be coarsely classified based on several
criteria: 1) target gate set; 2) algorithmic approach; and 3) solution distinguishability.

\parah{Target Gate Set} Some algorithms target gates likely
to be used only when fault tolerant quantum computing materializes. Examples include
synthesis of z-rotation unitaries with
Clifford+V approximation~\cite{Ross15} or Clifford+T
gates~\cite{KMM16,KSV02,Paetznick2014}.
While these efforts propelled the field of synthesis, they are not 
used on NISQ devices, which offer a different gate set
(e.g. $U_3, R_x, R_z,CNOT$ and M\o lmer-S\o rensen all-to-all).
Several~\cite{raban,synthcsd,ionsynth,davis2019heuristics}  
algorithms, discussed below target these gates directly. From our perspective,
since QFAST is composable and can invoke any synthesizer for instantiation, the existence of these algorithms indicates
that QFAST is portable across gate sets. 

\comment{ For example, the z-rotation unitaries can be
synthesized with Clifford+V approximation~\cite{Ross15} or  with Clifford+T gates~\cite{KMM16}. The set of Pauli, Hadamard, Phase, CNOT 
gates form what is known as the Clifford group gates. When augmented with the T gate defined
as 
\[
T = \left(
\begin{array}{cc}
1 & 0 \\
0 & \zeta_8 
\end{array}
\right), \ \ \mbox{where} \ \ \zeta_8 = e^{i\pi/4},
\]
the gate set is universal.
t lead to better complexity $\mathcal{O}(\log^{1.75}(1/\epsilon))$ compared 
to the SK Algorithm.  \comment{$\mathcal{O}(\log(1/\epsilon))$ T-count scaling.}
}  

\parah{Algorithmic Approaches} Most  earlier attempts inspired by
Solovay Kitaev use a recursive (or divide-an-conquer) 
formulation. More recent search based approaches are illustrated by the
Meet-in-the-Middle~\cite{MIM13}  algorithm.
Several  approaches \cite{23gates,householderQ} use techniques from linear algebra for
unitary/tensor decomposition, but there are open questions as to the suitability
for hardware implementation because  algorithms are expressed 
in terms of row and column updates of a matrix rather than in terms of qubits.

The state-of-the-art upper bounds on circuit depth are provided by
techniques~\cite{synthcsd,raban} that use Cosine-Sine
decomposition. The Cosine-Sine decomposition was first
used by~\cite{tucci}  for compilation purposes. In practice,
commercial compilers ubiquitously deploy only 
KAK decompositions for two qubit unitaries.
Khaneja and Glaser have applied the KAK Decomposition to more than
just 2-qubit systems \cite{khaneja2000cartan}.
For a 3-qubit system, it originally required 64 CNOTs \cite{vartiainen2004efficient}, which was later reduced to 40 CNOTs \cite{vatan2004realization}. We have shown above that this can be beat by any of the three synthesis tools tested in this work.
UniversalQ is
an exponent evaluated in this paper.
The basic formulation of these techniques is topology
independent.
The published approaches are hard to extend to different qubit gate
sets.

\comment{
There are not many studies published about synthesis of qutrit based
circuits and qutrit gate sets.~\cite{qtsynth} describes a method
using Givens rotations and Householder decomposition. As techniques
for qubit based systems using a similar approach have been
proposed~\cite{23gates}, they may allow an easier combination of
qutrit and qubit based synthesis. }

Several techniques~\cite{ionsynth,qaqc,davis2019heuristics} use numerical optimization
and report results for systems with at most four qubits. They
describe the single qubit gates in their variational/continuous representation and
use optimizers and search to find a gate decomposition and
instantiation. From these, we compare directly against~\cite{davis2019heuristics} which
is the only published optimal and topology-aware technique. For our
purposes, all these techniques seem to solve a combinatorial number of
hard (low distance) optimization problems. We expect QFAST to scale
better while providing comparable results. Furthermore, due to its
composability, we can directly leverage any of these implementations.

Topology awareness is important for synthesis algorithms, with
opposing trends. Most formulations assume all-to-all connectivity.
Specializing for topology in linear algebra
decomposition techniques seems to increase circuit depth by
rather large constants, ~\cite{synthcsd} mention a factor of nine,
improved by~\cite{raban} to $4\times$.  Specializing for topology in
search and numerical optimization techniques seems to reduce circuit
depth and Davis et al~\cite{davis2019heuristics} report up to  $4\times$  reductions.
We expect QFAST to behave like the latter.

\parah{Solution Distinguishability} 
Synthesis algorithms are classified as exact or approximate based on
distinguishability.  This is a subtle classification criteria, as most
algorithms can be viewed as either.  For example,~\cite{MIM13}
proposed a divide-and-conquer algorithm called Meet-in-the-Middle
(MIM). Designed for exact circuit synthesis, the algorithm
may also be used to construct an $\epsilon$-approximate circuit. The results seem to indicate that the algorithm failed
to synthesize a three qubit QFT circuit. 

Furthermore, on NISQ devices, the target gate set of the algorithm
(e.g. T gate) may
be itself implemented as an approximation when using native gates.

We classify our approach as approximate since we accept solutions at a small distance from the original
unitary. In a sense, when algorithms move from design to
implementation, all become approximate due to numerical
floating point errors.

\comment{
It allows one to search for
circuits of depth $l$ by only generating circuits of depth at most $\lceil l/2 \rceil$ at the complexity of $\mathcal{O}(|\mathcal{V}_{n,\mathcal{G}}|^{\lceil l/2\rceil}\log |\mathcal{V}_{n,\mathcal{G}}|^{\lceil l/2 \rceil})$, 
where $\mathcal{V}_{n,\mathcal{G}}$ denotes the set of unitaries for depth-one 
$n$-qubit circuit. The MIM algorithm is flexible and allows weights to be
added to the gate set to account for the possibility that some gates,
such as those that do not belong to the Clifford group, may be more expensive
to implement. It also allows ancillas to be used in the synthesis.  The 
algorithm uses a number of heuristics to prune the search tree.  Although
it was originally designed for exact circuit synthesis, the algorithm
may also be used to construct an $\epsilon$-approximate circuit.
}

%% file: sections/conc.tex
\section{Conclusion}

We have presented a quantum synthesis algorithm designed to produce short circuits and scale well in practice. The evaluation on depth optimal circuits, as well as circuits generated by domain  generators (VQE, TFIM) indicates that while not optimal, QFAST can significantly reduce the depth of circuits used in practice by domain scientists. This reduction is beyond the capabilities of other existing synthesis tools or optimizing compilers.
This bodes well for the future adoption of synthesis for algorithm discovery or circuit optimization during the NISQ era and beyond.

%% file: sections/psuedocode.tex
\section{Appendix: Pseudocode}

\begin{algorithm}
\caption{Algorithm}
\textbf{Input:} $U_t \in \mathcal{C}^{2^n\times 2^n}, K$ \newline
\textbf{Output:} $\text{P}$ \newline 
\textbf{Variables:} $U_t$ target unitary, $K$ the native synthesis tool, $P$ quantum program \newline
\textbf{Ensure:} $\Delta({\rm\it Compose}(P), U_t) \leq {\tt refinement\_distance}$
\begin{algorithmic}[1]
\State $k \gets  \text{native\_block\_size}(K)$
\State $A, L_F \gets \text{Decomposition}(U_t, k)$
\State $(P^{(i)})_{i=1}^d \gets \text{Instantiation}(A, K)$
\State $P \gets \text{Recombination}((P)_0^i, L_F)$
\Return $P$
\end{algorithmic}
\end{algorithm}

\begin{algorithm}
\caption{Decomposition}
\textbf{Input:} $U_t \in \mathcal{C}^{2^n\times 2^n}, k$ \newline
\textbf{Output:} $A = (\vec{\alpha}^{(i)})_{i=1}^d, L_f = (Q^{(i)})_{i=1}^d$ \newline 
\textbf{Variables:} $A$ list of gate's function values, $L_f$ list of fixed locations \newline
\textbf{Ensure:} $\Delta(\prod_{i=1}^d{F^{\otimes n}_m(\vec{\alpha}^{(i)},Q^{(i)})}, U_t) \leq {\tt refinement\_distance}$
\begin{algorithmic}[1]
\State blocks $\gets \{(U_t, \{1..n\})\}$
\While{$\exists b \in \text{blocks } s.t. \text{ sizeof}(b) > k$}
\State new\_blocks $\gets \{\}$
\ForAll{$b \in \text{blocks}$}
\State $m \gets \text{decomposition\_size}(b)$
\State $A, L \gets \text{exploration}(\text{fst}(b),m)$
\State $L_f \gets \text{fix\_locations}(L)$
\State $A \gets \text{refinement}(\text{fst}(b), m, A, L_f)$
\State $(U^{(i)})_{i=1}^{d^`} \gets \text{convert\_to\_unitary}(A)$
\State $(Q^{(i)})_{i=1}^{d^`} \gets \text{compose\_locations}(snd(b), L_f)$
\State new\_blocks $\gets \text{zip}((U^{(i)})_{i=1}^{d^`}, (Q^{(i)})_{i=1}^{d^`})$
\EndFor
\State blocks $\gets \text{new\_blocks}$
\EndWhile
\end{algorithmic}
\end{algorithm}

\begin{algorithm}
\caption{Exploration}
\textbf{Input:} $U_t \in \mathcal{C}^{2^n\times 2^n}, m$ \newline
\textbf{Output:} $A = (\vec{\alpha}^{(i)})_{i=1}^d, L = (\vec{l}^{(i)})_{i=1}^d$ \newline 
\textbf{Variables:} $A$ list of gate's function values, $L$ list of gate's location values \newline
\textbf{Ensure:} $\Delta(\prod_{i=1}^d{G^{\otimes n}_m(\vec{\alpha}^{(i)},\vec{l}^{(i)})}, U_t) \leq {\tt exploration\_distance}$ 
\begin{algorithmic}[1]
\State $d\gets0$
\Comment{Initialize empty circuit}
\State $A\gets()$
\State $L\gets()$
\While{True}
\State $d\gets d+1$
\Comment{Add another layer}
\State $A, L\gets \text{add\_layer}(A,L))$
\While{True}
\Comment{Minimize distance until}
\State $\text{loss}\gets\Delta(\prod_{i=1}^d{G^{\otimes n}_m(\vec{\alpha}^{(i)},\vec{l}^{(i)})}, U_t)$
\Comment{success or plateau}
\State $A, L \gets \text{Minimizer(loss)}$
\If{ $\text{loss}\leq{\tt exploration\_distance}$}
\State \Return A, L
\ElsIf{plateau}
\State Break
\EndIf
\EndWhile
\EndWhile
\end{algorithmic}
\end{algorithm}

\begin{algorithm}
\caption{Refinement}
\textbf{Input:} $U_t \in \mathcal{C}^{2^n\times 2^n}, m, A, L_f$ \newline
\textbf{Output:} $A = (\vec{\alpha}^{(i)})_{i=1}^d$ \newline 
\textbf{Variables:} $A$ list of gate's function values, $L_f$ list of fixed locations \newline
\textbf{Ensure:} $\Delta(\prod_{i=1}^d{F^{\otimes n}_m(\vec{\alpha}^{(i)},Q^{(i)})}, U_t) \leq {\tt refinement\_distance}$ 
\begin{algorithmic}[1]
\While{True}
\Comment{Minimize distance until}
\State $\text{loss}\gets\Delta(\prod_{i=1}^d{F^{\otimes n}_m(\vec{\alpha}^{(i)},Q^{(i)})}, U_t)$
\Comment{success or plateau}
\State $A \gets \text{Minimizer(loss)}$
\If{ $\text{loss}\leq{\tt refinement\_distance}$ \text{or} plateau}
\State \Return A
\EndIf
\EndWhile
\end{algorithmic}
\end{algorithm}

\begin{algorithm}
\caption{add\_layer}
\textbf{Input:} $A, L, n, m$ \newline
\textbf{Output:} $A, L$
\begin{algorithmic}[1]
\State $\vec{\alpha} \gets \{0\}^{2^m}$
\State $\vec{l} \gets \{0\}^{\binom{n}{m}}$
\State $A \gets append(A, \vec{\alpha})$
\If{$ \text{depth}(A) \equiv 0 (\text{mod } 2)$}
\State $L \gets append(L, \text{first\_half}(\vec{l}))$
\Else
\State $L \gets append(L, \text{second\_half}(\vec{l}))$
\EndIf
\end{algorithmic}
\end{algorithm}